# Improving FPGA resilience through Partial Dynamic Reconfiguration

José Luís Nunes
University of Coimbra/CISUC
Coimbra, Portugal
jnunes@dei.uc.pt

*Abstract*—This paper explores advances in reconfiguration properties of SRAM-based FPGAs, namely Partial Dynamic Reconfiguration, to improve the resilience of critical systems that take advantage of this technology. Commercial of-the-shelf state-of-the-art FPGA devices use SRAM cells for the configuration memory, which allow an increase in both performance and capacity. The fast access times and unlimited number of writes of this technology, reduces reconfiguration delays and extends the device lifetime but, at the same time, makes them more sensitive to radiation effects, in the form of Single Event Upsets. To overcome this limitation, manufacturers have proposed a few fault tolerant approaches, which rely on space/time redundancy and configuration memory content recovery – *scrubbing*. In this paper, we first present radiation effects on these devices and investigate the applicability of the most commonly used fault tolerant approaches, and then propose an approach to improve FPGA resilience, through the use of a less intrusive failure prediction *scrubbing*. It is expected that this approach relieves the system designer from dependability concerns and reduces both time intrusiveness and overall power consumption.

*Keywords*—dependability; failure prediction, real-time; embedded systems; FPGA; SEU

## I. INTRODUCTION

Field Programmable Gate Arrays (FPGA) were first introduced as prototyping platforms or as low-cost replacements in applications with low volume Integrated Circuits (IC) requirements. These devices are now increasingly being used for production in engineering areas, such as industrial, automotive and space, due to their flexibility and ability for fast in the field reconfiguration. Recent advances in FPGAs, take advantage of fast memory cells (SRAM) for storing configuration data, which allow shorter access times and an unlimited number of reconfigurations, when compared to flash-based devices. However, this memory technology is known to be very sensitive to radiation effects, manifested in the form of Single Event Upsets (SEU), thus limiting its applicability in the just mentioned areas. Therefore, some form of fault tolerance is mandatory to improve the resilience of these devices, and allow its use in critical systems.

Manufacturers have been studying the radiation effects on these devices for a long time [1], and have thus introduced a couple of radiation-hardened space-grade devices. These Radiation-Hardened-By-Process (RHBP) devices present severe limitations, in both cost and performance, when compared with Commercial Off-The-Shelf (COTS) devices. This led technology companies to start using less resilient COTS devices for deploying critical systems in harsh operating environments, and adding some form of fault tolerance.

To overcome SEU effects in SRAM-based FPGA configuration memory, one of the major manufactures, Xilinx, have: i) introduced error detection and correction mechanisms in their line of devices – frame Single Error Correction and Double Error Detection (SEC-DED) and device wide Cyclic Redundant Checks (CRC); ii) proposed to take advantage of spatial redundancy – using their X-TMR tool [2]; and iii) allowed configuration memory content to be refreshed (*scrubbing*). Although significantly improving device resilience, their overheads, in both device area usage and time delays, makes them unsuitable for remotely located critical real-time systems. Research community have also proposed alternative approaches [3] to error detection and correction, by means of partial reconfiguration (addressing just the faulty region), but most of them need external hardware and target very specific devices.

With the introduction of Partial Dynamic Reconfiguration (PDR) functionality by Xilinx, to support the time multiplexing of FPGA resources, it is now possible to keep a subset of the device running (*static module*) while other parts (*dynamic modules*) are being reconfigured. The implementation of an Internal Configuration and Access Port (ICAP), running at much higher speeds than the external configuration ports (*SelectMap*, JATG, etc.), opens the possibility to trigger the partial reconfiguration from the inside of the FPGA, at a much faster and predictable time, and therefore support critical real-time systems.

This work investigates the feasibility of using partial dynamic reconfiguration to improve the resilience of FPGA-based production systems, with strict timing requirements and constraints on device area usage. To accomplish this we propose an approach on how to continuously adapt the *scrubbing* parameters, using device health information, from both FPGA embedded sensors, and data from external sources. Through the use of failure prediction, we expect to: i) reduce the *scrubbing* mechanism time intrusiveness, by activating it only when needed; ii) improve overall power consumption, by activating the *scrubbing* less frequently; and iii) add an adaptive layer of system dependability, to cope with dynamic environments and hardware aging.

The paper starts with an introduction, where we identify the problem and introduce a few suitable technologies to address it. Following, in Section II, we describe the FPGA internals, the failure modes and the fault model. In Section III, we present a brief overview of current fault tolerance approaches and dependability assessment tools. Section IV presents the background for the on-going and future work that is presented in Section V, where we propose a few improvements over some well known Fault Tolerant (FT) approaches, and discuss the roadmap. Finally, in Section VI we present the conclusions of this work.

## II. FPGA FAULTS

Every SRAM-based FPGA is composed by a matrix of Configurable Logic Blocks (CLB), connected by routing resources, with additional storage and input/output elements. Every device has a Configuration Memory (CM) for storing the user implemented design/data, and control bits. At power-up, a *bitstream* file is loaded from the external permanent storage, usually a flash memory, to the internal configuration memory. The *bitstream* file is generated by an Hardware Definition Language (HDL) tool.

Radiation induced SEU in these devices may produce a large number of Single Event Effects (SEE), which generate one or more *bitflips* in the internal FPGA storage elements. Although having a common source, their effects could be very distinct, depending on the affected memory component. FPGA storage elements can be divided in five main groups: 1) Configuration Memory, which has the largest number of storage bits and is used to store the user design; 2) Block Memory (BRAM), used to store the design state; 3) Distributed Memory, which is part of specific CLBs, where the Lookup Table Elements (LUT) can be used as shift-registers or distributed RAM; 4) Flip-Flops, that are present in all CLBs and are used to store the design state; and 5) Internal Device Control and State, composed by a set of bits that are critical to the correct functioning of the device [4][5].

Configuration memory is the largest FPGA memory element and is used to store the user design [6]. While upsets in other memory elements of the FPGA may crash the device completely or corrupt the design/system state, these upsets, although transient, may produce a permanent change in the implemented functional design (logic and routing), and ultimately a system failure. According to [7], these faults represent more than 80% of the total faults affecting FPGAs. Also, due to the large amount of Programmable Interconnect Points (PIP) needed to interface all the CLB matrix, a large percentage of configuration memory bits used in the implementation are thus non-sensitive (faults that affect these bits do not produce any changes in the design).

Depending on the number of affected cells per upsets, we may classify them as Single-Bit Errors (SBE) or Multiple-Bit Errors (MBE). Although the frequency of MBEs is increasing, due to the high-level of integration in recent devices, SBEs are much more frequent, with an approximate ratio of 20:1 [8]. The shape of MBEs usually follows a circular pattern around the radiation impact centre cell, and may spread over multiple frames (a frame is a single bit column that spans the entire configuration memory vertically [4]).

Our focus is on transient and intermittent faults that affect FPGA configuration memory, which may produce permanent errors in the functional logic and routing resources. This is justified because some FPGA components, like Flip-Flops (FF), are inherently more resilient to SEU, and also because memory locations that store user data (BRAM) are actively protected by Error Correction Codes (ECC). This is not the case with the configuration memory cells that implement the user design logic, were the computed frame ECC may be easily accessed, but single-bit errors are not automatically corrected, relaying this decision to the system designer.

## III. FPGA FAULT TOLERANCE

Xilinx, one of the two major FPGA manufacturers, has been introducing several fault tolerant technologies for SRAM-based devices. Mechanisms such as: *readback* CRC checks during boot-up and throughout system execution; and frame ECC bits for single-bit error correction and double-bit error detection, are available to the system designer to improve system resilience.

### A. Existing Approaches

*Scrubbing*, the process of refreshing configuration memory data, can be implemented in different ways: blindly full/partial *scrubbing* to restore the initial configuration memory contents; using *SEC-DED* codes and PDR to locate and correct the error; and *Readback and Compare,* which continuously monitors the configuration memory and reload/reconfigure, when necessary. The main drawbacks of *scrubbing* are: i) delays introduced in accessing configuration memory; ii) the size of the minimum addressable block, a frame, that includes a few hundreds bits; and iii) the need to use external circuitry in devices that do not support PDR.

*Spatial Redundancy* is the most used approach to mitigate the errors induced by SEUs. A popular implementation of this type of redundancy is Triple Modular Redundancy (TMR), which triplicates the implemented system and the associated voters in the feedback loops. When applied to FPGAs, TMR should be combined with some form of partial *scrubbing* to restore the fault tolerance level, by reconfiguring the affected TMR domain. Examples of commercially available tools that automatically generate a full/partial TMR protected system from a non-protected one are: X-TMR, STMR, and BL-TMR. The main drawback of this approach is the FPGA resource overhead (3+ times the size of the unprotected system), and the need to recover the faulty TMR domains to restore the previous fault tolerance level. This is not feasible in cost sensitive application and remote location deployment, like space, where weight is at a premium.

*Temporal Redundancy* is able to mask transient errors, but special care should be taken in order to handle permanent errors. Device resource overhead is small but the introduced delays limit its applicability to non real-time systems.

### B. Dependability Assessment

To assess the dependability properties of an FPGA-based embedded system and the fault tolerance mechanisms in place, one possibility is to inject faults in the system and monitor its behaviour. Depending on the technology used for the injection of errors in the FPGA configuration memory, the methods are grouped in hardware-based fault injection (FI) and software-based FI. The high maintenance costs of the hardware-based FI facilities and the need to prepare the device die surface before each FI campaign make it unaffordable to most institutions. To overcome this limitation, some researchers took advantage of PDR properties to simulate SEE in the FPGA configuration memory, assuming a *bitflip* fault model. On the other side, software-based FI has higher precision, better controllability and lower costs, but pose a few concerns related to the representativeness of the injected faults.

There are only a couple of FPGA software-based fault injectors, mainly from academia research projects, and these are targeted at now obsolete FPGA device families (caused by major changes in the FPGA internal design after the introduction of each new family). All of these fault-injectors use some form of *bitstream* instrumentation, either before system startup (offline) or during system execution (online).

- *FLIPPER and FLIPPER2* [9]. Targeted at Xilinx Virtex-2 and Virtex-4, respectively, they use partial reconfiguration to inject single and multiple *bitflips* in configuration memory. FLIPPERs mimic the radiation experiments and allow to identify design sensitive bits and evaluate the SEU sensitivity of an ASIC prototype implemented in an FPGA.
- *FT-UNSHADES* [10]. This fault injector was created to study the effects of radiation-induced faults in ASICs, assuming a *bitflip* fault model. This system is able to inject *bitflips* in FFs, through PDR, on a Xilinx FPGA. The FPGA implements both the "faulty system" and the "fault-free system" and compares their output to inform the host computer that a failure occurred.
- *SEU Simulator* [11]. It uses PDR to corrupt the FPGA frames while the system is running. It is composed of three FPGAs, two running the user circuit in parallel (golden design and design under test) and a third one responsible for real-time output comparison. The simulator can be used to classify the user design configuration bits as being sensitive or non-sensitive.
- *SEU Controller Macro* [12]. Developed by Xilinx as a reference design for a Virtex-5 FPGA development board, this module is able to simulate hardware faults, and corrupt configuration memory bits, in order to assess error detection and correction mechanisms. This macro takes advantage of PDR to inject single and double adjacent *bitflips* in configuration memory cells.

IV. CURRENT WORK

As preliminary results of this work, in [13] we have described our initial thoughts on how to improve the resilience of FPGA-based systems by taking advantage of PDR. Our main goal was reducing the FPGA fault tolerance mechanism footprint while increasing, or at least providing the same level of fault tolerance achieved with current approaches. At the same time, we introduce the idea of proactive adaptation of the fault tolerance mechanisms to the evolving requirements of the operation environment.

Using a development board from Digilent – XUPV5, with a Xilinx Virtex-5 chip that supports PDR, we were able to build a System-On-Chip (SoC) based on a single FPGA. This setup allowed us to acquire deep knowledge of the dynamic reconfiguration process, and identify the steps ahead. In this setup the recovery of the faulty system was performed by a Microblaze soft core that access FPGA configuration memory through the ICAP interface, for faster reconfiguration speeds.

In [14], using the setup described above, we detailed the dynamic properties of real-time control systems that could benefit from using PDR, as the underlying technology for implementing fault recovery mechanisms. Considering the controlled system time constant and the size of the implemented controller in the FPGA, it may be feasible to reduce space redundancy to a minimum and simply recover the module during the available idle time in between control loops. The main questions addressed in this contribution were: How fast can we reconfigure a Reconfigurable Partition (RP) of a specific size? How does this interfere with the control loop and the controller performance?

To assess the dependability of FPGA-based systems, we needed a tool to perform fault injection at FPGA configuration memory level. Thus, in [15], we evaluated the Xilinx SEU Controller Macro (SEU-CM) as a first basis for a fault injection tool. This tool uses PDR, through the ICAP interface, to recover single-bit errors from configuration memory, in a *read-modify-write* approach. As it relies on a single SEC-DED code per configuration memory frame (1312 bits in a Virtex-5 device), it only allows the detection of single and double-bit errors, and provides the identification and correction of single-bit errors. It also includes the possibility of changing the configuration memory contents on the fly, using the same mechanism. Thus, system designers can take advantage of it, as a primitive tool to assess the fault tolerant mechanism in use.

The results of the SEU-CM evaluation showed that, although not initially developed to support fault injection, it could be used to inject errors in FPGA configuration memory cells, with some limitation:

- Low efficiency, as it randomly injects faults in unused FPGA location. It is thus necessary to clearly identify the area of the FPGA where the Device Under Test (DUT) is located, and target only this area.
- Low observability of the DUT at low level due to the little amount of information available from Xilinx on the *bitstream* internal format. One could, at the system design level, attach probes to the RP interfaces in order to monitor the I/O and perform data logging.
- Limited fault model, which only considers single or double adjacent *bitflips*. While single frame multiple *bitflip* injection can easily be achieved by changing SEU-CM VHDL code, simultaneously injecting multiple *bitflips* in adjacent frames cannot be done instantaneously, as each frame is *read-modified-written* at a time. Thus simulation of radiation-induced multiple-bit upsets is compromised.
- Lack of automation, to extend the fault injection campaigns and collect enough data to attain statistical significance.

Driven by the need to assess the dependability properties of the proposed fault tolerance approaches, and the inexistence of a reference fault injector for SRAM-based FPGA user designs, we developed our own fault injector. This tool, Fault Injector for Reconfigurable Embedded Devices (FIRED) [16], uses the PDR error injection approach of the SEU-CM as a starting point, and extends its capabilities by: i) defining regions of interest for the FI campaign; ii) extending the fault model to include multiple-bit errors that span beyond a single frame; iii) improving observability by attaching probes to the RP interfaces and logging I/O data to an external SQL database; and iv) precisely defining each injected fault (location, trigger, type, pattern, etc.).

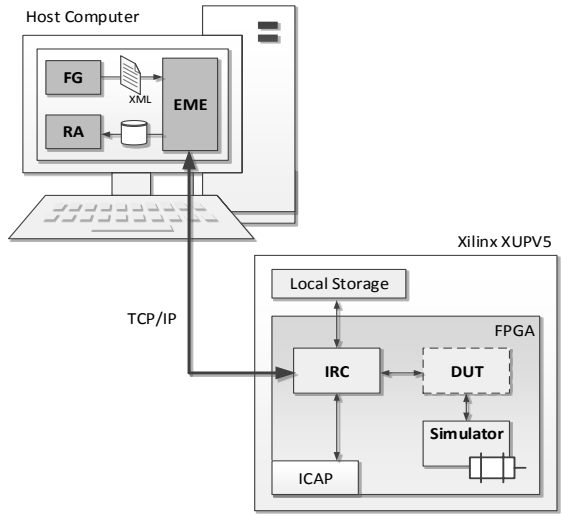

Fig. 1. Implemented fault injector (FIRED) hardware and modules, which uses PDR for error insertion in FPGA configuration memory.

In Figure 1, we show the hardware components and modules that compose FIRED. On the hardware side, is uses a PC to define and control the experiments, and an FPGA *devboard* to implement the low level error injection and data collection modules. The PC executes the Fault Generator (FG), the Experiment Management Environment (EME), and finally the Result Analysis (RA) components. On the *devboard*, the Injection Runtime Controller (IRC) is responsible for: i) introducing errors in the FPGA configuration memory, using ICAP; ii) collecting input/output DUT data for offline analysis (saved to the Local Storage); and iii) controlling the DUT and the workload. The communication between the *devboad* and the PC is accomplished through a TCP/IP connection.

The fault injector IRC is implemented on the unused cells of the FPGA, side-by-side with the DUT. The later is confined to an RP (depicted in Figure 1 by a dashed line), which, by the intrinsic parallelism of the FPGA, provides the needed isolation from FIRED components.

Following, to evaluate the radiation effects in the FPGA configuration memory and the device failure modes, we implemented a PID-based cruise control system in VHDL, to be used as a *testbed* (see Figure 2). This controller complies with the general FIRED DUT interface: 32-bit input, 32-bit output, CLK line, Enable and Reset signals. The main goals were: i) to study the effects of *bitflips* in the controller (in this specific case a PI controller); ii) to identify the more sensitive areas of the FPGA; and iii) to verify the feasibility of blindly *scrubbing*, and see how it copes with the strict timing requirements of real-time systems.

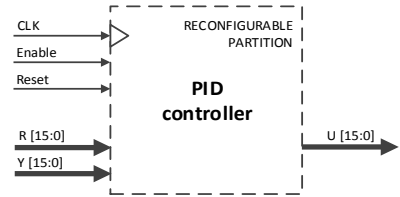

Fig. 2. PID-based cruise controller module with the I/O signals.

This controller can be connected to an external DC motor with shaft rotation encoder, or to a VHDL simulated system. The simulated system allows the user to speedup the system dependability assessment and provides a completely deterministic system.

In [17], we used FIRED to perform an extensive fault injection campaign in the PID-based cruise control system, and characterise the cumulative SEU effects in real-time systems. Two important results obtained were related to the fault latency and fault interaction.

In this FI campaign we injected 10 single-bit faults in accumulation throughout each experiment, with 42 seconds duration. The workload was composed by a periodic square wave with two different settings for the desired cruise control speed. The outcome of each experiment was compared with a *goldrun* (fault free experiment) to detect a system failure. Any deviation from the desired output was classified as a failure (which accounted for 28.4% of all experiments).

The root cause analysis showed that the vast majority of failures were caused by a single fault (99.4%). The remaining 0.6% of the failures was caused by interacting faults, which wouldn't occur if they were injected isolated. Analysing the latency of the single-fault failures, we noticed that more that half have high error latency. These results allow us to think that fault tolerant approaches based on *scrubbing* could be successfully applied to these types of systems, if we tackle a few problems, namely: power consumption overhead, system state recovery, and time/space FT mechanism intrusivness.

With a better understanding of FPGA resource allocation it may be possible to avoid blindly *scrubbing* unused frames of the configuration memory and, as there are a large amount of bits in the used frames where a *bitflip* won't have any effect on the implemented logic, even avoid some of the used frames when non-sensitive bits are affected. The main question here is whether to refresh a couple of frames that map to a high-level logic unit (e.g. the proportional part of a PID controller), or all the user design. This information is mandatory to recover the system during control loop idle time, in critical real-time system that don't show high error latencies as the one evaluated.

The short amount of information provided by the manufacturers about the mapping between the high-level HDL

descripting of the system and the low-level *bitstream* data generated by their tools (Xilinx only provides information about the LUT location on their application notes), makes it difficult to understand the effects of *bitstream* errors in the controller logic. There are a few 3rd-party tools [18] [19] that tried to reverse engineer the *bitstream* contents and generate an XDL file, with the logic resource allocation and routing resources, but they are targeted at very specific devices and the results are far from being complete. This limitation led us to instead monitor the interfaces of the DUT as an approach to detect a system failure.

The main goal of this work is not related to the error detection and reactive system recovery, but to improve the fault tolerance by means of preventive and proactive fault tolerance approaches. The error detections and correction in such systems is by itself a full topic of research, and the specificities and the limitations of the FPGA error detection mechanisms impose a prohibitive time overhead that could affect the functionality of the implemented design.

## V. WORKPLAN

Critical real-time systems have very stringent timing requirements, which may not allow the designer to follow the guidelines from manufacturers [8]. The use of simple periodic, full or partial, *scrubbing* in such systems may impose a huge overhead in power consumption, intrusiveness, and lack the adaption needed by changing environments.

Our goal is to provide the system designer with an Intellectual Property (IP) core that could be used to improve the resilience of FPGA-base production system, without the need to evaluate and decide on the *scrubbing* approaches proposed by manufacturers. This solution will allow the system to have the same level of CM latent error removal, while adapting the *scrubbing* frequency to the device status and the operating environment properties.

The workplan can be divided in three steps: i) identify and correlate device and operating environment sensor data, with the device health status; ii) implement failure prediction *scrubbing* using the previously identified data sources with machine learning algorithms; iii) validate this approach, comparing it with the periodic blindly *scrubbing*, by means of a fault injection campaign. In the following subsections, we detail the work involved in each step and how we plan to approach it.

Finally, we shall also explore the concept of dependable adaptive systems which, instead of reducing their computation to the minimum when entering an harsh environment to limit interferences and avoid system failures, change their modular structure to make them more resilient. This will allow reducing device usage to a minimum throughout time, which will have a positive impact on power consumption.

### A. Device and Environment Sensor Data Sources

In order to fine-tune the *scrubbing* subsystem, it is necessary to identify which data sources provide a better insight of the device condition and environment properties. This is accomplished by executing extensive FI experiments and data correlation.

Internal data coming from FPGA embedded sensors, can provide inputs about the device status. In the specific case of the Virtex-5 device, it is possible to know the working temperature and all the DC-DC converters output voltage. If we could correlate any fluctuation in the converters output voltage or the expected temperature rise, due to a high overload or a short circuit (when a SEU affects a routing resources), we may as well predict an eminent failure of the protected module.

It is also possible to implement additional probes inside the FPGA configuration memory, that may provide valuable information about the switching delays in CLB internal components, or that are able to capture very short duration glitches in their outputs [20].

External information may come from different sources, such as a web service, that could provide information about solar radiation bursts and environmental condition (temperature, humidity, pressure, etc.), or even external sensors connected to the FPGA (instant power consumption, etc.).

### B. Failure Prediction Scrubbing

To take advantage of all available data, we propose a system architecture based on an additional component, which we named *fpScrub*, implemented in the FPGA programmable logic. This component is responsible for triggering the partial reconfiguration (*scrubbing*) of the *Protected module*, when needed. He gathers information from different sources, both internal and external to the FPGA.

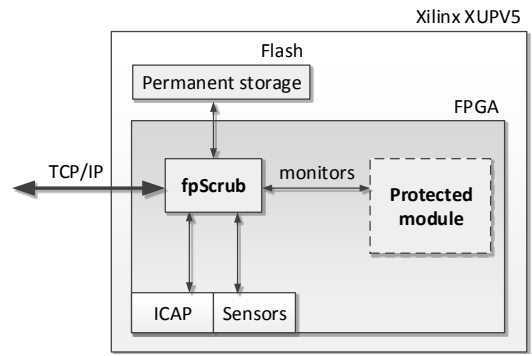

Fig. 3. Failure predition *scrubbing* implementation in a Xilinx XUPV5 (Virtex-5) development board, showing the *fpScrub* component and his interfaces (dashed lines limit the reconfigurable partition).

In Figure 3, we present the system architecture, with the main component, *fpScrub*, monitoring the *Protected module* I/O and connected to: i) ICAP, through a PLB bus, for reconfiguration purposes; ii) internal FPGA sensors, to monitor device health; and iii) external data sources, through an Ethernet interface, to gather operating environment data.

In specific situations, where the range of the *Protected module* inputs and outputs are bounded and known in advance, the *fpScrub* component could continuously monitor his interfaces to detect an upcoming problem.

### C. System Validation and Improvements

The proposed approach will be validated using fault injection, and the results compared with the periodic blindly

*scrubbing*. The outcomes of these FI experiments are expected to show improvements, in both temporal intrusiveness and device power consumption, when applied to systems with high error latency. On the contrary, in systems with low error latency, the overhead of the *fpScrub* will be significant as its *scrubbing* frequency approaches the periodic blindly *scrubbing*.

Afterwards, if we are able to identify the device elements where the user designs reside and understand the mapping to the high-level representation of the system, we could increase the *scrubbing* granularity, by splitting the *bitstream* in smaller chunks. This allows the proposed approach to address more complex systems, that otherwise wouldn't be supported. In situations where the *fpScrub* is unable to completely refresh the *Protected module*, in between two control loops, it tries to refresh as many as it can. During the recovering period the user design will be running in a degraded mode. As soon as the last piece is recovered, the user design is in an identical state as it was just after the FPGA boot-up sequence. In order to do this, we need to assure that the recovery won't affect the system state. Otherwise we need to consider some form of *checkpointing*.

If we wish to extend the fault model to include permanent faults in FPGA logic cells, we need to address the problem of static routing in RPs (sometimes HDL tools use internal RP routing resources to implement a shorter path between two external components). Otherwise it won't be possible to dynamically relocate the system to other FPGA cells in the presence of hardware defects. The presence of intermittent faults may be a good indicator of the occurrence of these phenomena, such as hardware aging and manufacturing defects.

## VI. CONCLUSIONS

This paper described the PhD research work, the current state and the future plans in exploiting Partial Dynamic Reconfiguration of FPGA devices to improve FPGA-based production systems dependability.

In the current work we started by analysing the PDR capabilities of current devices, for improving the fault tolerance of SRAM-based FPGAs. Following we evaluated the SEU Controller Macro, as a free available tool for fault injection. In order to automate the full FI campaign, we developed our own fault injector for FPGA-based embedded devices – FIRED. We finished with a study on the effects of cumulative faults in a PID-based cruise control system, which showed high error latency and the evidence of interacting faults.

Giving continuity of this work, in the workplan we identify the final steps to implement a new approach to improve FPGA resilience, through an adaptive *scrubbing*. This takes into account device health information from embedded sensors, together with external environmental data, to trigger *scrubbing* only when needed. When compared to the periodic blindly *scrubbing*, we expect it reduce both the temporal intrusiveness and the power consumption, while providing the same level of fault tolerance in static environments. We also expect it to improve FPGA-based system resilience in evolving operating environments and under device hardware aging, as it will provide the system with an adapive layer of fault tolerance.